\documentclass[a4paper]{jpconf}
\usepackage{graphicx}
\begin{document}
\title{Third Harmonic Flow of Charged Particles in Au+Au Collisions at $\sqrt {s_{NN}} = 200$ GeV}

\author{Yadav Pandit for the STAR Collaboration}

\address{Dept. of Physics, Kent State University, Kent, OH 44242}

\ead{ypandit@kent.edu}

\begin{abstract}
 In this  proceedings, we report measurements of the third harmonic coefficient of the azimuthal
anisotropy, $v_{3}$, known as triangular flow. The analysis is for charged particles near midrapidity in Au+Au collisions at $\sqrt {s_{NN}} $ = 200 GeV, based on data from the STAR experiment at the Relativistic Heavy Ion Collider.  Triangular flow as a function of centrality, pseudorapidity and transverse momentum are reported using various methods, including a study of the signal for particle pairs as a function of their pseudorapidity separation. 
%%; analyses of this type can help distinguish among different sources of non flow correlations.  
Results are compared with other experiments and model predictions.
\end{abstract}

\section{Introduction}

The study of azimuthal anisotropy, based on Fourier coefficients, is widely recognized as an important tool to probe the hot, dense matter created in heavy ion collisions~\cite{methodPaper}.
  %%%%%%%%%%%%%%%%%%%%%%%%%%%%%%%%

\begin{equation}
v_n = \langle \cos n( \phi-\Psi_R ) \rangle
\end{equation}
%%%%%%%%%%%%%%%%%%%%%%%%%%%%%%%%%%%%%%%
where $\phi$ denotes the azimuthal angle of the outgoing particles, $\Psi_R$ is the orientation of the reaction plane and $n$ denotes the harmonic.

The first harmonic coefficient $v_1$, called directed flow, and the second harmonic coefficient $v_2$, called elliptic flow, have been extensively studied both experimentally~\cite{v1andv2} and theoretically, while higher even-order harmonics have also garnered some attention~\cite{v4v6}.  In contrast, odd harmonics of order three and above were overlooked until recently, because in a picture with smooth 
initial overlap geometry, it had been assumed that higher-order odd harmonics are required to be zero by symmetry. Now it is realized that event-by-event fluctuations break this symmetry~\cite{geoFluct1, derik, riseFall}. As a consequence, higher-order odd harmonics carry valuable information about the initial state of the colliding system~\cite{Mishra, geoFluct2, geoFluct3, transpov3, 
v3_4D-hydro, hydrov3, v3-AMPT}.  The third harmonic coefficient --- related to a collective motion known as triangular flow --- is thus a new tool to study initial-state fluctuations, and the subsequent evolution of the collision system.  Theoretical studies suggest that $v_{3}$ is more sensitive to viscous effects than $v_2$ because finer details of the higher harmonics are smoothed more by viscosity. We present measurements of $v_{3}$ vs. the pseudorapidity separation $\Delta \eta$ = $\eta_{i} - \eta_{j}$ between the two particles, fit with narrow and wide Gaussians, and present results for the wide  Gaussian as a function of $p_{T}$ and centrality. 

\section{Methods and Analysis}
Azimuthal correlations not related to the initial geometry and reaction plane orientation which can arise from resonances, jets, strings, quantum statistics effects, final state interactions (particularly Coulomb effects), and momentum conservation are refereed as  non-flow correlations. Two-particle correlations which are sensitive to non-flow correlations, and multiparticle correlations which are less sensitive to non-flow correlations, along with event plane methods have been extensively used in elliptic flow measurements for the past couple of decades.  These methods can be extended to measure triangular flow. The event plane method and the two-particle cumulant method are used in the present study.    

\subsection{Event Plane Methods}
%%%%%%%%%%%%%%%%%%%%%   Fig Gaussians %%%%%%%%%%%%%%%%%%%%%%%%%%
\begin{figure}
\resizebox{0.75\textwidth}{!}{
  \includegraphics{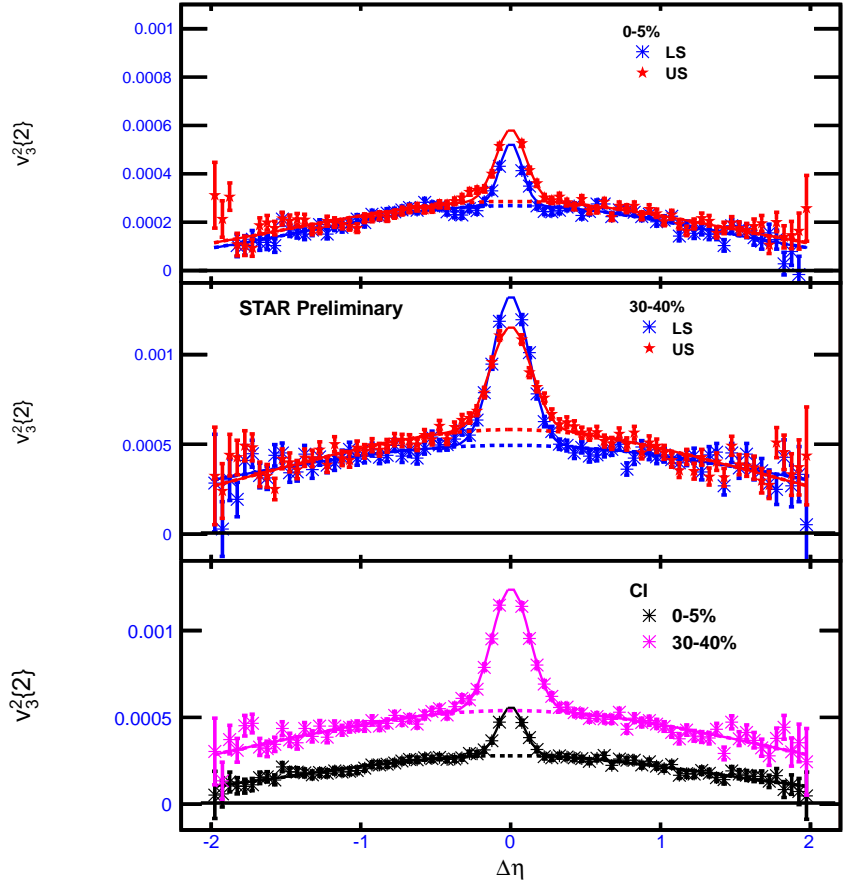}} 
\caption { $v_{3}^{2}\{2, \Delta \eta \}$ vs. $\Delta \eta $ for charged hadrons within two centrality intervals in 200 GeV Au+Au collisions. Data are fit with narrow and wide Gaussians. Like sign (LS), 
unlike sign (US), and charge independent (CI) cases are shown.}
\label{figuregauss}      
\end{figure}
  
   In the standard event plane method~\cite{methodPaper} for $v_3$, we reconstruct a third-harmonic event plane from Time Projection Chamber (TPC ($|\eta|<1.0 $)) tracks and also from Forward TPC( $2.5< |\eta|<4.0$)  tracks. The event plane vector $Q_{n}$ and the
event plane angle $\Psi_{3}$ from the third  harmonic of the particle azimuthal distribution are defined by the equations
\begin{equation} \label{Q3x} Q_3\cos (3\Psi_3)\ =\ Q_{3x}\ =\
\sum_{i}w_i\cos (3\phi_i),
\end{equation}
\begin{equation} \label{Q3y} Q_3\sin (3\Psi_3)\ =\ Q_{3y}\ = \
\sum_{i}w_i\sin (3\phi_i),
\end{equation}
\begin{equation} \label{Psi} \Psi_3\ =\
 \tan^{-1} ( \frac{Q_{3y}}{Q_{3x}})/3,
\end{equation}
In event plane calculations, tracks have a weighting factor $w = p_T$ in units of GeV$/c$ for $p_{T} < 2$ GeV $/c$, and w = 2 for $p_{T} \geq $ 2 GeV$/c$.  This helps to maximize the event plane resolution and also helps to reduce the contribution from jets in event plane calculations. Although the STAR detector has good azimuthal symmetry, small acceptance effects in the calculation of the event plane azimuth were removed by the method of shifting~\cite{ShiftMethod}. When using the TPC event plane, we used the $\eta$ sub-event method with an additional $\eta$ gap of $\pm$ 0.05.
\begin{equation}
v_{3}\{\rm {EtaSub}\}\   = \frac{\langle \cos 3(\phi_{\pm} - \Psi_{3, \eta_{\mp}})\rangle}{\sqrt{\langle \cos 3(\Psi_{3, \eta_{+}} - \Psi_{3, \eta_{-}})\rangle}}
\label{etasub}
\end{equation}
 This avoids self-correlations because the particles and the event plane are in opposite hemispheres. When using the FTPCs, we used the full event plane from both FTPCs~\cite{methodPaper}. 
 
\begin{equation}
v_{3}\{\rm {FTPC}\}\   = \frac{\langle \cos 3 (\phi - \Psi_{3FTPC})\rangle}{ C \sqrt{\langle \cos 3 (\Psi_{3, FTPC\,\eta_{+}} - \Psi_{3, FTPC\,\eta_{-}})\rangle}}
\label{fullFTPCep}
\end{equation}
This introduces a large $\eta$ gap between the particles and the event plane. Since there is no overlap between the coverage of the TPC and FTPCs, there is no possibility of self-correlation when using the FTPC event plane.

%%%%%%%%%%%%%
\subsection{Two-Particle Correlation Method}

\begin{figure}
\centering
\includegraphics[width=4in]{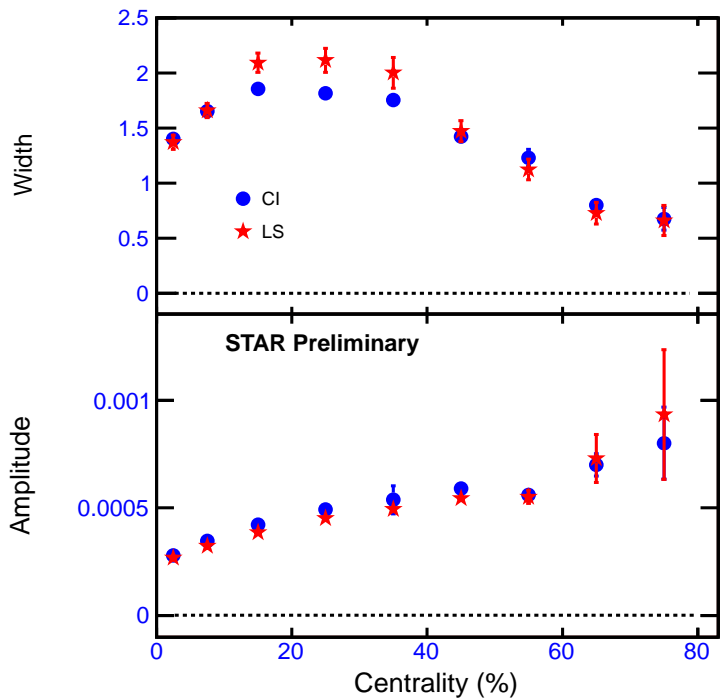}
\caption{ Width and amplitude of the wide Gaussian as a function of centrality  for charge-independent (CI) and like-sign (LS) particles. The plotted errors are statistical.}
\label{gaussianCen}
\end{figure}

We study $\langle \cos 3(\phi_j-\phi_i) \rangle_{i \neq j} $ vs $ \Delta \eta$ for two particles with indices $i$ and $j$  to understand the  $\Delta \eta$  dependence of the  triangular flow signal  and to  distinguish among different sources of  non-flow correlations.  This distribution of $\langle \cos 3(\phi_j-\phi_i) \rangle_{i \neq j}$  vs. $ \Delta \eta $   can be well described by wide and narrow Gaussian peaks as shown in Fig.~\ref{figuregauss} for two centrality intervals. The narrow Gaussian is identified with short range non-flow correlations like  Bose-Einstein  correlations,  resonance decay, Coulomb interactions, and effects from track merging. The narrow peak disappears above $p_{T} \geq 0.8$ GeV/$c$  so it is unlikely to be from jet correlations. The wide Gaussian represented by $v_{3}^{2}\{2, \Delta \eta \}$ is  the signal of interest  and its fit parameters are used to calculate $v_{3}^2\{2\}$  as a  function of centrality and transverse momentum.  For  $v_{3}$ integrated over $p_{T}$ and $\eta$ we have  

\begin{figure}
\centering
\includegraphics[width=4in]{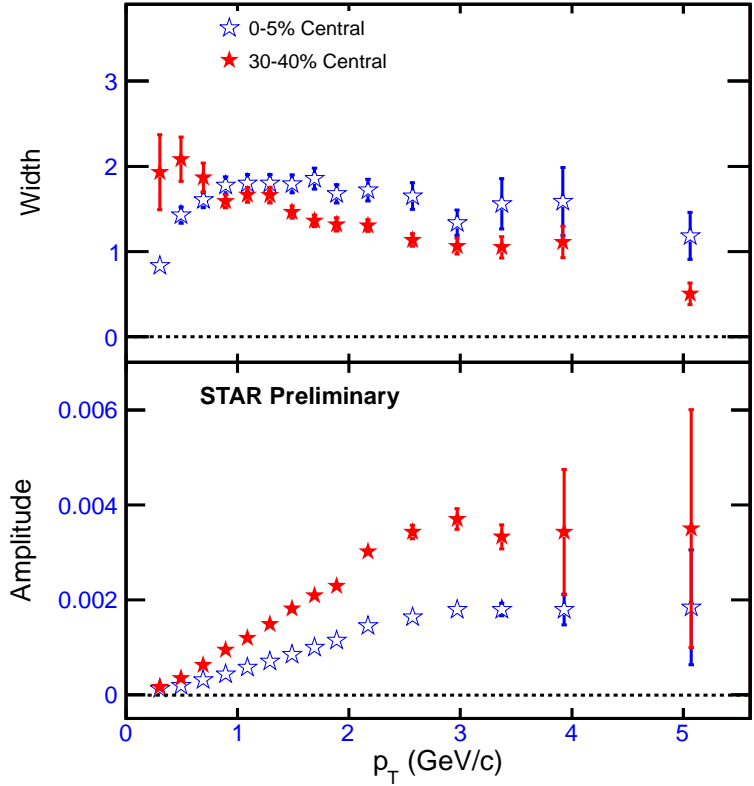}
\caption{The upper panel shows the width, and the lower panel shows the amplitude of the wide Gaussian, as a function of transverse momentum, for the most central (0--5\%) and for mid-central (30--40\%) collisions. The plotted errors are statistical.}
\label{gaussianPt}
\end{figure}

 %%%%%%%%%%%%%%%%%%%%%%%%% Eqs. %%%%%%%%%%%%%%%%%%%%%%%%%%%%%%%%%%
\begin{equation} 
  v_{3}^{2}\{2\}     =   \frac {\int^b_a v_{3}^{2}\{2, \Delta \eta \}  W d(\Delta \eta )} {\int^b_a W d(\Delta \eta)} ,
\label{eq:int} 
\end{equation}
 %%%%%%%%%%%%%%%%%%%%%%%%% Eqs. %%%%%%%%%%%%%%%%%%%%%%%%%%%%%%%%%%% 
where $W$ equals $dN/d(\Delta \eta) $ when weighted with the number of particle pairs, or is set to 1 for unit weight. The quantity $v_{3}\{2\}(p_T)$ can be obtained from  the scalar product~\cite{ScalarProduct} relation
\begin{equation} 
 v_{3}\{2\}(p_T)   =   \frac{\mbox{$\langle \cos 3(\phi_j-\phi_i) \rangle_{i \neq j} $}} {\sqrt{\left< v_{3}^{2}\{2\} \right>}}  ,
\label{eq:dif} 
\end{equation} 
where the $j^{\rm th}$ particle is selected from the $p_{T}$ bin of interest.
%%%%%%%%%%%%%
\section{Results and Discussion}
%%%%%%%%%%%%%%%%%%%%%   Fig centrality %%%%%%%%%%%%%%%%%%%%%%%%%%
\begin{figure*}
\centering
\resizebox{0.65\textwidth}{!}{
  \includegraphics{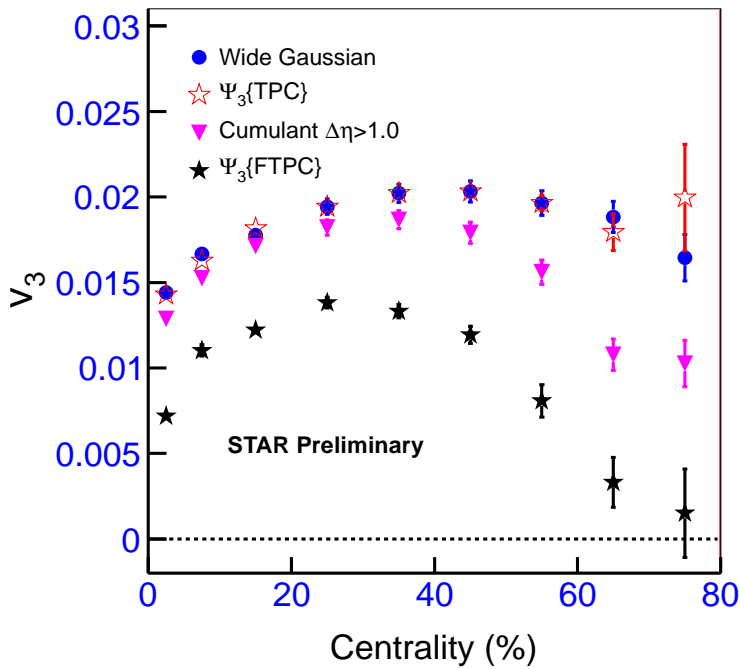} }
\caption {The third harmonic coefficient as a function of centrality from different methods of measurement for Au+Au collisions at $\sqrt {s_{NN}} = 200$ GeV, with track selections $0.15 < p_T < 2.0$ GeV/$c$ and $-1.0 < \eta < 1.0$.}
\label{fig:cent}      
\end{figure*}
Figure~\ref{gaussianCen} shows the width and amplitude of the wide Gaussian as a function of centrality. We observe that the width peaks in mid-central collisions whereas the amplitude increases from central to peripheral collisions.
Figure~\ref{gaussianPt} shows the width and amplitude of the wide Gaussian as a function of $p_{T}$ for centralities 0--5\% and 30--40\%. Above $p_{T} \geq$  0.8 GeV/$c$, the distribution can be described by a single wide Gaussian. The amplitude increases with $p_T$ and then saturates around $p_{T} =$ 3 GeV/$c$. The $p_{T}$ dependence  of the width seems to depend on centrality, with the 0--5\% most central data showing first an increase and then a gradual decrease, while at 30--40\% centrality, the data appear to gradually decrease as a function of  $p_{T}$. 
Figure~\ref{fig:cent} shows the centrality dependence for $p_{T}$-integrated  $v_3$ from several different analyses: $v_{3}\{2\}$ from Eq.~\ref{eq:int} and Fig.~\ref{figuregauss} for the wide Gaussian, $v_{3}\{\rm{TPC}\}$ measured with third harmonic event plane reconstructed in the TPC, two-particle cumulants with a minimum pseudorapidity separation between particles of one unit,  and $v_{3}\{\rm{FTPC}\}$ where $v_{3}$ is measured relative to the third harmonic event plane reconstructed  in the FTPCs. 
\begin{figure}
\centering
\resizebox{0.65\textwidth}{!}{
  \includegraphics{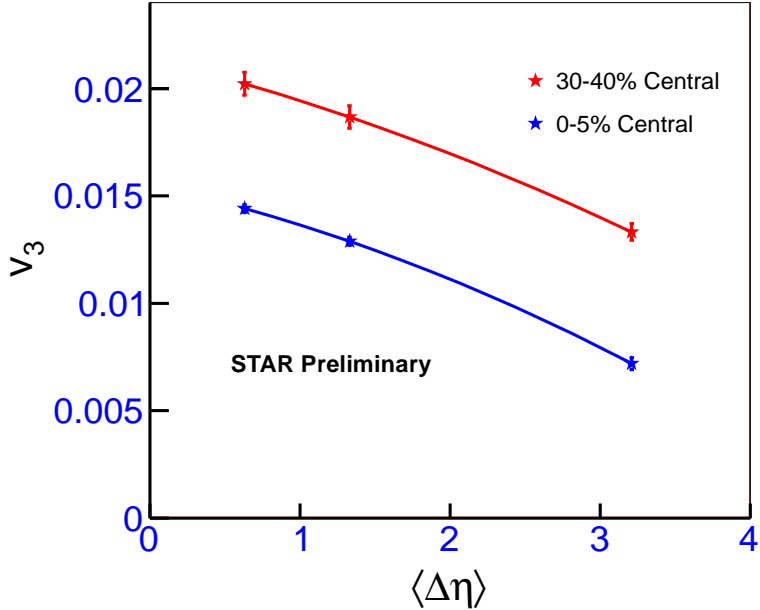} }
\caption {The third harmonic coefficient as a function of mean $\Delta \eta $. The two points at $\langle\Delta\eta\rangle = 0.63$ are from the methods using the TPC with $ |\eta| < 1$.  The points at 1.33 are from the cumulant results with $|\Delta \eta| > 1$. The points at 3.21 are from correlations with the FTPC event plane.} 
\label{fig:v3DeltaEta}      
\end{figure}
%%%%%%%%%%%%%%%%%%%%%%%%%%%%%%%%%%%%%%
\begin{figure}
\centering
\resizebox{0.65\textwidth}{!}{
  \includegraphics{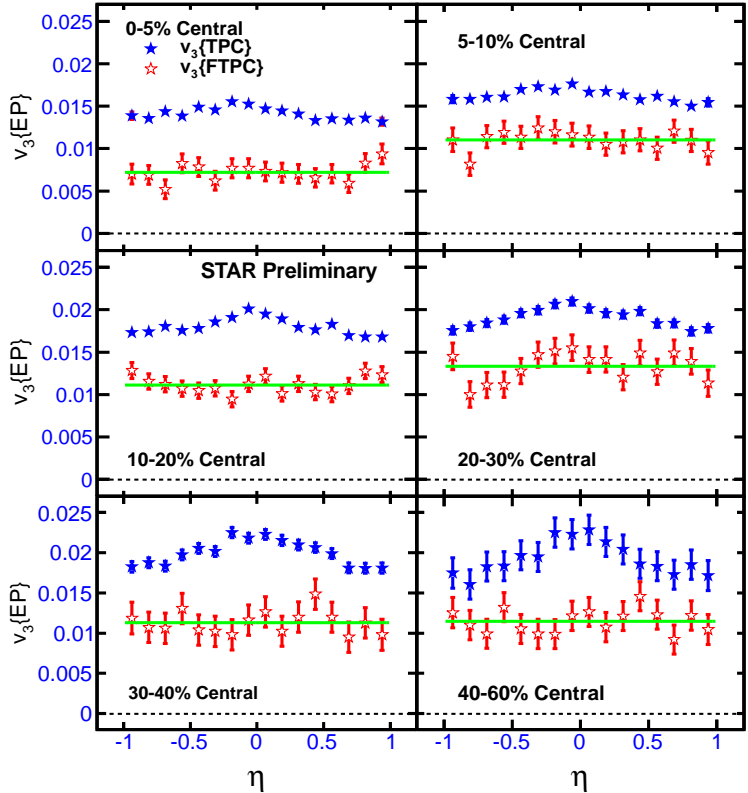} }
\caption {The third harmonic coefficient as a function of $\eta$ for different centralities for Au+Au collisions at $\sqrt{s_{NN}} = 200$ GeV, with track selection in the TPC of $0.15 <  p_T <  2.0$ GeV/$c$ and $ |\eta| <  1.0$. Results are shown for the event plane constructed both in the TPC and in the FTPCs. The horizontal lines are fits to the FTPC results.} 
\label{fig:eta}      
\end{figure}
%%%%%%%%%%%%%%%%%%%%%   Fig PHENIX-STAR  Pt dependence %%%%%%%%%%%%%%%%%%%%%%%%%%
\begin{figure}
\centering
\resizebox{0.85 \textwidth}{!}{
  \includegraphics{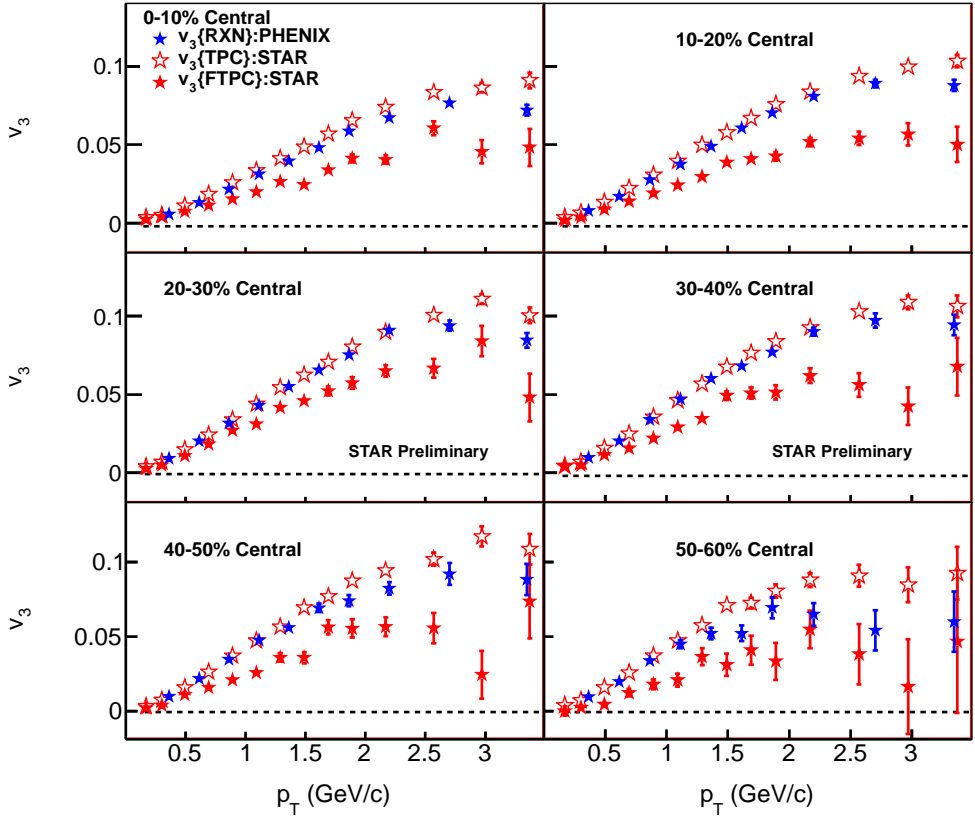} }
\caption {The third harmonic coefficient as a function of $p_T$ for both event plane methods for different centralities compared to the results from the PHENIX experiment~\cite{Adare:2011tg}. PHENIX requires $|\eta| < 0.35$ while STAR requires $|\eta| < 1.0$.  In the case of STAR results from the TPC, the mean $\Delta\eta$ was 0.63, while in the case of the FTPC event plane, the average $\Delta\eta$ was 3.21. The PHENIX results used the event plane from their \rm{RXN} detector at an intermediate $\eta$ of $1.0 <  \eta < 2.8$.} 
\label{fig:phenix_starFTPC}      
\end{figure}
%%%%%%%%%%%%%%%%%%%%%   Model comparisons  %%%%%%%%%%%%%%%%%%%%%
\begin{figure*}
\centering
\resizebox{0.85 \textwidth}{!}{
  \includegraphics{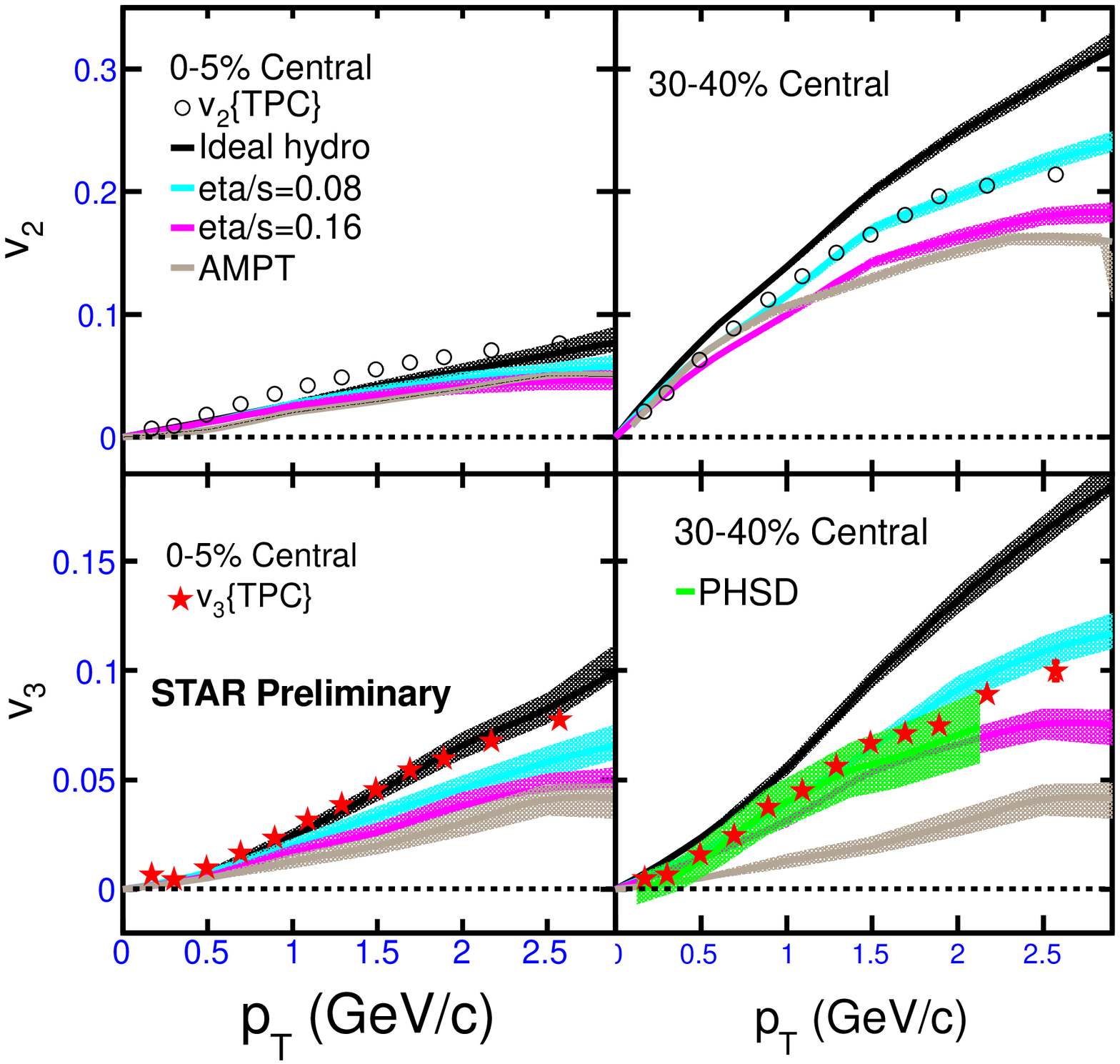}}
\caption{ $v_2$ (top) and $v_3$ (bottom) for Au+Au collisions at $\sqrt{s_{NN}} = 200$ GeV in 0--5\% and 30--40\% central collisions as a function of transverse momentum, compared with ideal~\cite{transpov3} and viscous~\cite{hydrov3} hydro, AMPT transport~\cite{v3-AMPT}, and Parton-Hadron String Dynamics~\cite{PHSD} models. The STAR $v_2$ values (top) come from Ref.~\cite{v4v6}.}  
\label{fig:models} 
\end{figure*}

%%%%%%%%%%%%%
Clearly the various analysis methods for $v_{3}$ differ greatly in Fig.~\ref{fig:cent}. For the wide Gaussian and the TPC event plane, the results are similar, suggesting that the narrow Gaussian is not very important. When a large $\Delta \eta$ is required, the results decrease, especially for peripheral collisions. The  variation between results shown in Fig.~\ref{fig:cent} is caused by the $\Delta \eta$ dependence as shown in Fig.~\ref{fig:v3DeltaEta}. Non-flow effects, which are assumed to be largest when $\Delta \eta$ is small, are expected to contribute less to the methods sampling a larger $\langle |\eta| \rangle$. This hypothesis is consistent with  previous studies of elliptic flow, based on two-particle correlations, where the corresponding wide Gaussian was ascribed to mini-jet correlations~\cite{minijet}. The decrease with $\Delta \eta$ of $v_n^2\{2\}$ may arise from decreasing non-flow or from decreasing initial-state density fluctuations; it is argued that those fluctuations decrease with increasing $\Delta \eta$ separation~\cite{decoherence}. Since $v_3$ depends so strongly on the $\eta$ separation of the particles, one must always quote $\Delta \eta$ for each $v_3$ measurement, and compare results to models with the same $\Delta \eta$. If $v_{3}^{2}\{2\}$ is related to the initial eccentricity fluctuations, then the reduction of $v_{3}^{2}\{2\}$ at large $\Delta \eta $ would presumably require a decrease of the initial state fluctuations at large rapidity separations. Recent work has found such a decoherence effect with a hadron and parton cascade model~\cite{decoherence}.

The non-flow contributions due to short-range correlations are effectively suppressed by using the wide Gaussian or by the $\eta$-gap method. We studied the influence of different pseudorapidity gaps. However, Fig.~\ref{fig:v3DeltaEta} shows that we did not find a pattern where $v_3$ stabilized at a constant value for large $\Delta \eta$. We compared like and unlike charge-sign combinations, since they have different contributions from resonance decays, fluctuations, and final state interactions. The large discrepancy between the methods may have its origin in the $\Delta \eta$ dependence of the fluctuations, which seem to decrease with increasing $\Delta \eta$ separation. Thus it is not clear if one should extrapolate to large $\Delta \eta$ to avoid non-flow, or to small $\Delta \eta$ to measure all the fluctuations~\cite{decoherence}.  More theoretical input is necessary to completely understand the $\Delta \eta $ dependence of this signal.
Figure~\ref{fig:eta} shows the $\eta$ dependence of $v_3$ using event plane methods. For particles in the TPC using the $\eta$ sub-event method, $v_3$ is somewhat peaked at mid-rapidity. With the event plane in the FTPCs, there is a large $\eta$ gap and $v_3$ is flat at all centralities. This implies that acceptance effects at the edges of the TPC are not important. Thus, even though a large $\Delta\eta$ means that one of the particles is probably at large $\eta$, this apparently does not strongly influence the $\Delta \eta$ dependence.

In Fig.~\ref{fig:phenix_starFTPC}, STAR results with the event plane in the TPC are very similar to those of PHENIX~\cite{Adare:2011tg}. This is surprising because the mean $\eta$ of their \rm{RXN} detector is larger than in the case of the sub-events in the STAR TPC. However, the STAR FTPC results are lower than the PHENIX results. This is expected, because the mean $\Delta \eta$ is considerably larger in the FTPC than in the PHENIX RXN detector.

Third harmonic flow coefficient has been studied in event-by-event ideal hydrodynamics~\cite{hydrov3} with MC Glauber initial conditions. These authors concluded that instead of averaged initial conditions, event-by-event calculations are necessary to compare with experimental data. The first prediction of $v_3$ with viscous hydro was Ref.~\cite{transpov3}.  In Fig.~\ref{fig:models}, $v_{2}\{TPC\}$ and $v_{3}\{TPC\}$ as a function of transverse momentum  obtained with TPC sub event plane method are compared with several models for 0--5\% and 30--40\% central collisions. The experimental results for the TPC sub-event plane method which may have small non flow contribution are shown because they eliminate the short-range correlations but yet have small $|\Delta \eta|$ like the theory calculation. The specific models are the viscous hydrodynamic model of Ref.~\cite{hydrov3}, where the ratio of viscosity to entropy is $\eta/s = $ 0.08 and 0.16, and the AMPT model~\cite{v3-AMPT}. AMPT model results are for $320 < N_{part}  <  360 $  and $ 80 < N_{part}  < 120$  from Ref ~\cite{geoFluct1} compared with 0-5\% and 30-40\% central data .  Predictions for $v_{3}$ from Parton-Hadron String Dynamics~\cite{PHSD} at 30--40\% centrality for $|\eta| <  0.5$ have been made by the sub-event method with the event planes at $1.0 <  |\eta| < 4.0$, and are also plotted in Fig.~\ref{fig:models} (lower right panel).

Elliptic flow($v_{2}\{TPC\}$)  results are mostly described by ideal hydrodynamics in the case of the most central collisions, and by $\eta/s = $ 0.08 in the case of mid-central collisions.  We find that the third harmonic coefficient( $v_{3}\{TPC\}$) results are also described by this model with similar viscosities. The PHSD model also agrees with data.

%%%%%%%%%%%%%%%%%%
\section{Summary}

We present STAR's measurements of third harmonic flow $(v_{3})$  of charged particles from Au+Au collisions at $\sqrt {s_{NN}} = 200$ GeV as a function 
of pseudorapidity, transverse momentum, and centrality.  We report result rom a two-particle method for particle pairs with an gap or fit with a wide Gaussian in pseudorapidity separation, as well as from the standard event-plane method with an event plane near midrapidity or at forward rapidity. Short-range correlations are eliminated either by an gap or by  eliminating  the narrow Gaussian in pseudorapidity separation. The measured values of $v_{3}$ 
continuously decrease as the mean pseudorapidity separation($\langle\Delta\eta\rangle$)  of the particles increase within the range observable by STAR.  It is not known whether this decrease is due to a decrease in non-flow correlations or due to a decrease in fluctuations.  We observe that $v_{3}$ increases with transverse momentum before it levels-off above 3 GeV$/c$, similar to the case of elliptic flow.  Our results are mostly described by hydrodynamic models with small viscosity with MC Glauber initial conditions.

\section{Acknowledgments}
%%%%%%%%%%%%%%%%%%%%%%%%%%%%%%%%

  We thank the RHIC Operations Group and RCF at BNL, the NERSC Center at LBNL and the Open
Science Grid consortium for providing resources and support. This work was supported in part by the
Offices of NP and HEP within the U.S. DOE Office of Science, the U.S. NSF.


\section{References}

\begin{thebibliography}{9}

\bibitem{methodPaper}
A. M. Poskanzer and S. A. Voloshin, Phys. Rev. C {\bf 58}, 1671 (1998).

\bibitem{v1andv2}
G. Agakishiev  {\it et al.} (STAR~Collaboration), Phys. Rev. C {\bf 85}, 014901(2012).
Y. Pandit (for the STAR Collaboration)  J. Phys. Conf. Ser.{\bf 316}, 012001 ( 2011).
 B. I. Abelev {\it et al.}  (STAR Collaboration), Phys. Rev. Lett. {\bf 101}, 252301 (2008).
 B. I. Abelev {\it et al.} (STAR Collaboration), Phys. Rev. C {\bf 81}, 044902 (2010).

\bibitem{v4v6}
P. F.~Kolb, Phys. Rev. C {\bf 68}, 031902(R) (2003);
J.~Adams {\it et al.} (STAR~Collaboration), Phys. Rev. Lett. {\bf 92}, 062301 (2004).

%J.~Adams {\it et al.} (STAR~Collaboration), Phys. Rev. C {\bf 72}, 014904 (2005).

\bibitem{geoFluct1}
B. Alver and G. Roland, Phys. Rev. C {\bf 81}, 054905 (2010).
%%this is first paper on v3 by alver

\bibitem{derik} 
D. Teaney and L. Yan, Phys. Rev. C{\bf 83}, 064904 (2011). 

\bibitem{riseFall}
P. Sorensen, B. Bolliet, A. Mocsy, Y. Pandit and N. Pruthi,  Phys. Lett. B {\bf 705} , 71( 2011).  

\bibitem{Mishra}
A. P. Mishra, R. K. Moohapatra, P. S. Saumia and A. M. Shrivastava, Phys. Rev. C {\bf 81}, 034903 (2010).

\bibitem{geoFluct2}
G.-Y. Qin, H. Petersen, S. A. Bass, and B. M\"uller, Phys. Rev. C {\bf 82} , 064903 (2010).  

\bibitem{geoFluct3}
J. L. Nagle and M. P. McCumber, Phys. Rev. C{\bf 83}, 044908 (2011).

\bibitem{transpov3}
B. H. Alver, C. Gombeaud, M. Luzum and J. Y. Ollitrault, Phys. Rev. C {\bf 82}, 034913 (2010).

\bibitem{v3_4D-hydro}
H. Petersen, G.-Y. Qin, S. A. Bass, and B. M\"uller, Phys. Rev. C {\bf 82}, 041901(R) (2010).

\bibitem{hydrov3}
B. Schenke, S. Y. Jeon, and C. Gale, Phys. Rev. Lett. {\bf 106}, 042301 (2011). 

\bibitem{v3-AMPT}
J. Xu and C. M. Ko, Phys. Rev. C {\bf 84} 014903 (2011).

\bibitem{ShiftMethod} 
J. Barrette {\it et al.} (E877 Collaboration), Phys. Rev. C {\bf 56}, 3254 (1997).

\bibitem{ScalarProduct} 
C. Alder {\it et al.} (STAR Collaboration), Phys. Rev. C {\bf 66}, 034904 (2002);
N. Borghini, P. M. Dinh, and J. Y. Ollitrault, Phys. Rev. C {\bf 63}, 054906 (2001). 


%\cite{Ollitrault:2009ie}
\bibitem{Ollitrault:2009ie} 
  J.~-Y.~Ollitrault, A.~M.~Poskanzer and S.~A.~Voloshin,
  %``Effect of flow fluctuations and nonflow on elliptic flow methods,''
  Phys.\ Rev.\ C {\bf 80}, 014904 (2009).
  %%[arXiv:0904.2315 [nucl-ex]].
  %%CITATION = ARXIV:0904.2315;%%

%\cite{Bozek:2012en}
\bibitem{Bozek:2012en} 
  P.~Bozek and W.~Broniowski,
  %``Charge conservation and the shape of the ridge of two-particle correlations in relativistic heavy-ion collisions,''
  arXiv:1204.3580 [nucl-th].
  %%CITATION = ARXIV:1204.3580;%%
  
  %\cite{Adare:2011tg}
\bibitem{Adare:2011tg} 
  A.~Adare {\it et al.}  (PHENIX Collaboration),
  %``Measurements of Higher-Order Flow Harmonics in Au+Au Collisions at $\sqrt{s_{NN}} = 200$ GeV,''
  Phys.\ Rev.\ Lett.\  {\bf 107}, 252301 (2011).
  %%[arXiv:1105.3928 [nucl-ex]].
  %%CITATION = ARXIV:1105.3928;%%
  
  \bibitem{whitePapers}
%   I.~Arsene {\it et al.}  (BRAHMS~Collaboration),    Nucl. Phys. A {\bf 757}, 1 (2005);
 %  B.~B.~Back {\it et al.} (PHOBOS~Collaboration),   Nucl. Phys. A {\bf 757}, 28 (2005);
   J.~Adams {\it et al.}  (STAR~Collaboration),    Nucl. Phys. A {\bf 757}, 102 (2005).
 %  K.~Adcox {\it et al.}  (PHENIX Collaboration),   Nucl. Phys. A {\bf 757}, 184 (2005).
  
%%Dipole assmetry 
\bibitem{PHSD}
 V. P. Konchakovski, E. L. Bratkovskaya, W. Cassing, V. D. Toneev, S. A. Voloshin, and V. Voronyuk, Phys. Rev. C {\bf 85}, 044922 (2012).

\bibitem{Cumulant}
A. Bilandzic, R. Snellings, and S. Voloshin, Phys. Rev. C {\bf 83}, 044913 (2011).

\bibitem{ollitro}
J. Y. Ollitrault, Phys. Rev. D {\bf 46}, 229 (1992).

\bibitem{STAR200GeV-v2}
 J. Adams {\it et al.} (STAR Collaboration), Phys. Rev. C {\bf 72}, 014904 (2005).

\bibitem{globerMC} 
  M. L. Miller, K. Reygers, S. J. Sanders and P. Steinberg, Ann. Rev. Nucl. Part. Sci. {\bf 57}, 205 (2007);
   B. Alver et al., Phys. Rev. C {\bf 77}, 014906 (2008).
 
 \bibitem{decoherence}
  K. Dusling, F. Gelis, T. Lappi and R. Venugopalan, Nucl. Phys. A {\bf 836}, 159 (2010).    
  
  \bibitem{cascadeModel}
  H. Petersen, C. Greiner, V. Bhattacharya and S. A. Bass, arXiv:1105.0340 [nucl-th].

 \bibitem{minijet}
   M. Daugherity (STAR Collaboration), J. Phys. G {\bf 35}, 104090 (2008); D. Kettler (STAR
Collaboration), PoS C ERP2010, 011 (2010); T. A. Trainor and D. T. Kettler, Phys. Rev.
C {\bf 83}, 034903 (2011).

\end{thebibliography}
\end{document}